\documentclass[
    ,final            
    ,numberedheadings 
  ]
  {aipproc}

\layoutstyle{6x9}

\begin{document}

\title{A \textsl{Suzaku} View of Cyclotron Line Sources and Candidates}



\author{K. Pottschmidt}{
address={CRESST and NASA Goddard Space Flight Center, Astrophysics
  Science Division, Code 661, Greenbelt, MD 20771, USA},
altaddress={Center for Space Science and Technology, University of
  Maryland Baltimore County, 1000 Hilltop Circle, Baltimore, MD 21250,
  USA}
}

\author{S. Suchy}{
address={University of California, San Diego, Center for Astrophysics
  and Space Sciences, 9500 Gilman Dr., La Jolla, CA 92093-0424, USA}
}

\author{E. Rivers}{
address={University of California, San Diego, Center for Astrophysics
  and Space Sciences, 9500 Gilman Dr., La Jolla, CA 92093-0424, USA}
}

\author{R. E. Rothschild}{
address={University of California, San Diego, Center for Astrophysics
  and Space Sciences, 9500 Gilman Dr., La Jolla, CA 92093-0424, USA}
}

\author{D. M. Marcu}{
address={CRESST and NASA Goddard Space Flight Center, Astrophysics
  Science Division, Code 661, Greenbelt, MD 20771, USA},
altaddress={Center for Space Science and Technology, University of
  Maryland Baltimore County, 1000 Hilltop Circle, Baltimore, MD 21250,
  USA}
}

\author{L. Barrag\'an}{
address={Dr. Karl Remeis-Observatory, Sternwartstr. 7, 96049 Bamberg,
  Germany},
altaddress={Erlangen Centre for Astroparticle Physics, University of
  Erlangen-Nuremberg, Erwin-Rommel-Strasse 1, 91058 Erlangen, Germany}
}

\author{M. K\"uhnel}{
address={Dr. Karl Remeis-Observatory, Sternwartstr. 7, 96049 Bamberg,
  Germany},
altaddress={Erlangen Centre for Astroparticle Physics, University of
  Erlangen-Nuremberg, Erwin-Rommel-Strasse 1, 91058 Erlangen, Germany}
}

\author{F. F\"urst}{
address={Dr. Karl Remeis-Observatory, Sternwartstr. 7, 96049 Bamberg,
  Germany},
altaddress={Erlangen Centre for Astroparticle Physics, University of
  Erlangen-Nuremberg, Erwin-Rommel-Strasse 1, 91058 Erlangen, Germany}
}

\author{F. Schwarm}{
address={Dr. Karl Remeis-Observatory, Sternwartstr. 7, 96049 Bamberg,
  Germany},
altaddress={Erlangen Centre for Astroparticle Physics, University of
  Erlangen-Nuremberg, Erwin-Rommel-Strasse 1, 91058 Erlangen, Germany}
}

\author{I. Kreykenbohm}{
address={Dr. Karl Remeis-Observatory, Sternwartstr. 7, 96049 Bamberg,
  Germany},
altaddress={Erlangen Centre for Astroparticle Physics, University of
  Erlangen-Nuremberg, Erwin-Rommel-Strasse 1, 91058 Erlangen, Germany}
}

\author{J. Wilms}{
address={Dr. Karl Remeis-Observatory, Sternwartstr. 7, 96049 Bamberg,
  Germany},
altaddress={Erlangen Centre for Astroparticle Physics, University of
  Erlangen-Nuremberg, Erwin-Rommel-Strasse 1, 91058 Erlangen, Germany}
}

\author{G. Sch\"onherr}{
address={Astrophysikalisches Institut Potsdam, 14482 Potsdam, Germany}
}

\author{I. Caballero}{
address={CEA SAclay, DSM/IRFU/SAp --UMR AIM (7158)
  CNRS/CEA/Univ.\ P.\ Diderot --F-91191, France}
}

\author{A. Camero-Arranz}{
address={Universities Space Research Association, Huntsville, AL
  35806, USA},
altaddress={Space Science Office, VP62, NASA/Marshall Space Flight
  Center, Hunstville, AL 35812, USA}
}

\author{A. Bodaghee}{
address={Space Sciences Laboratory, 7 Gauss Way, University of
  California, Berkeley, CA 94720, USA}
}

\author{V. Doroshenko}{
address={Institut f\"ur Astronomie und Astrophysik Astronomie, Sand 1,
  72076 T\"ubingen, Germany}
}

\author{D. Klochkov}{
address={Institut f\"ur Astronomie und Astrophysik Astronomie, Sand 1,
  72076 T\"ubingen, Germany}
}

\author{A. Santangelo}{
address={Institut f\"ur Astronomie und Astrophysik Astronomie, Sand 1,
  72076 T\"ubingen, Germany}
}

\author{R. Staubert}{
address={Institut f\"ur Astronomie und Astrophysik Astronomie, Sand 1,
  72076 T\"ubingen, Germany}
}

\author{P. Kretschmar}{
address={ESA-European Space Astronomy Centre, 28691 Villanueva de la
  Ca\~nada, Madrid, Spain}
}

\author{C. Wilson-Hodge}{
address={Space Science Office, VP62, NASA/Marshall Space Flight
  Center, Hunstville, AL 35812, USA}
}

\author{M. H. Finger}{
address={Universities Space Research Association, Huntsville, AL
  35806, USA},
altaddress={Space Science Office, VP62, NASA/Marshall Space Flight
  Center, Hunstville, AL 35812, USA}
}

\author{Y. Terada}{
address={Graduate School of Science and Engineering, Saitama
  University, 255 Simo-Ohkubo, Sakura-ku, Saitama City, Saitama
  338-8570, Japan}
}

\begin{abstract}
Seventeen accreting neutron star pulsars, mostly high mass X-ray
binaries with half of them Be-type transients, are known to exhibit
Cyclotron Resonance Scattering Features (CRSFs) in their X-ray
spectra, with characteristic line energies from 10 to 60\,keV. To date
about two thirds of them, plus a few similar systems without known
CRSFs, have been observed with \textsl{Suzaku}. We present an overview
of results from these observations, including the discovery of a CRSF
in the transient 1A\,1118$-$61 and pulse phase resolved spectroscopy
of GX~301$-$2. These observations allow for the determination of
cyclotron line parameters to an unprecedented degree of accuracy
within a moderate amount of observing time. This is important since
these parameters vary -- e.g., with orbital phase, pulse phase, or
luminosity -- depending on the geometry of the magnetic field of the
pulsar and the properties of the accretion column at the magnetic
poles. We briefly introduce a spectral model for CRSFs that is
currently being developed and that for the first time is based on
these physical properties. In addition to cyclotron line measurements,
selected highlights from the \textsl{Suzaku} analyses include dip and
flare studies, e.g., of 4U\,1907$+$09 and Vela~X-1, which show clumpy
wind effects (like partial absorption and/or a decrease in the mass
accretion rate supplied by the wind) and may also display
magnetospheric gating effects.
\end{abstract}

\maketitle


\section{Introduction}

\begin{figure}
\includegraphics[width=0.49\textwidth]{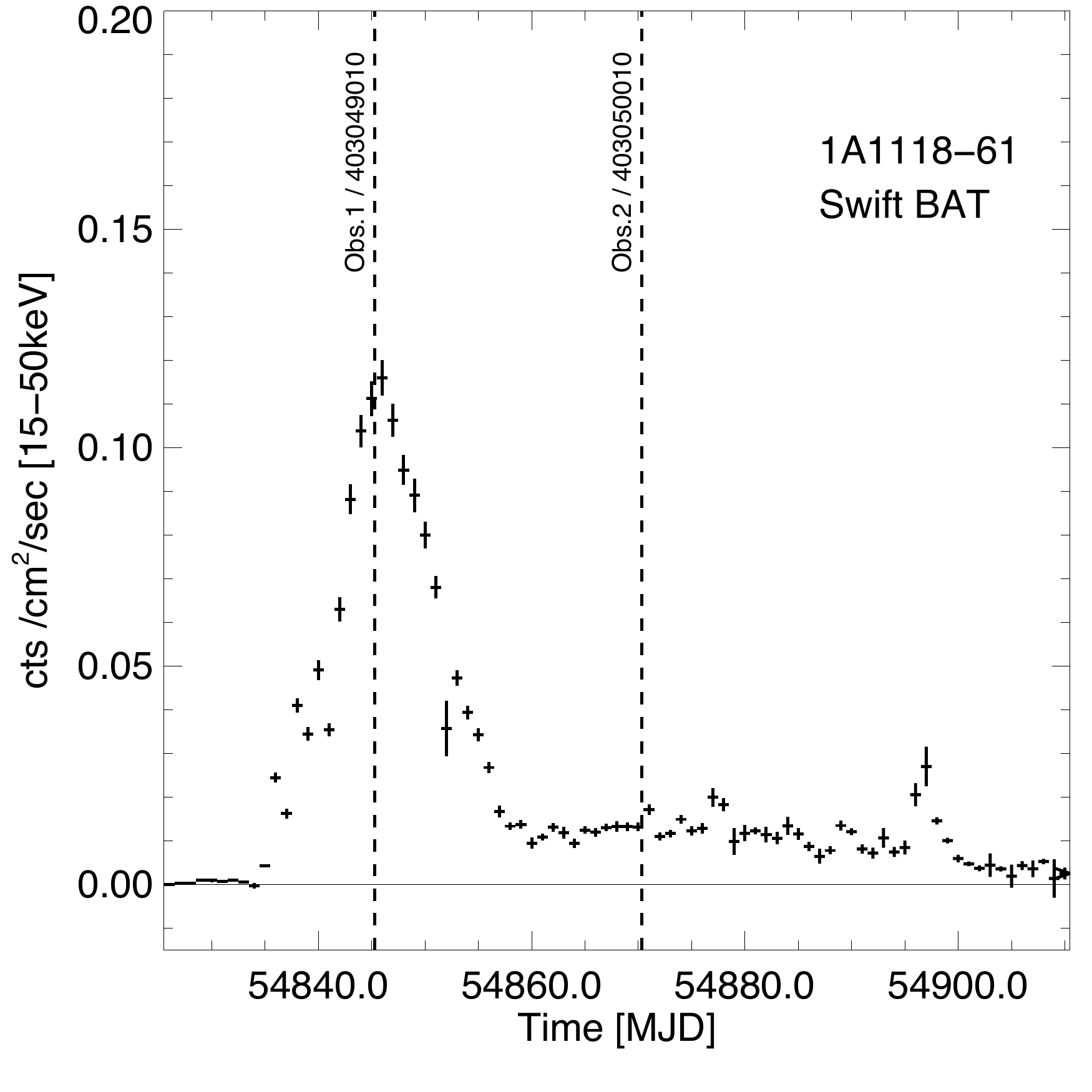}
\includegraphics[width=0.49\textwidth]{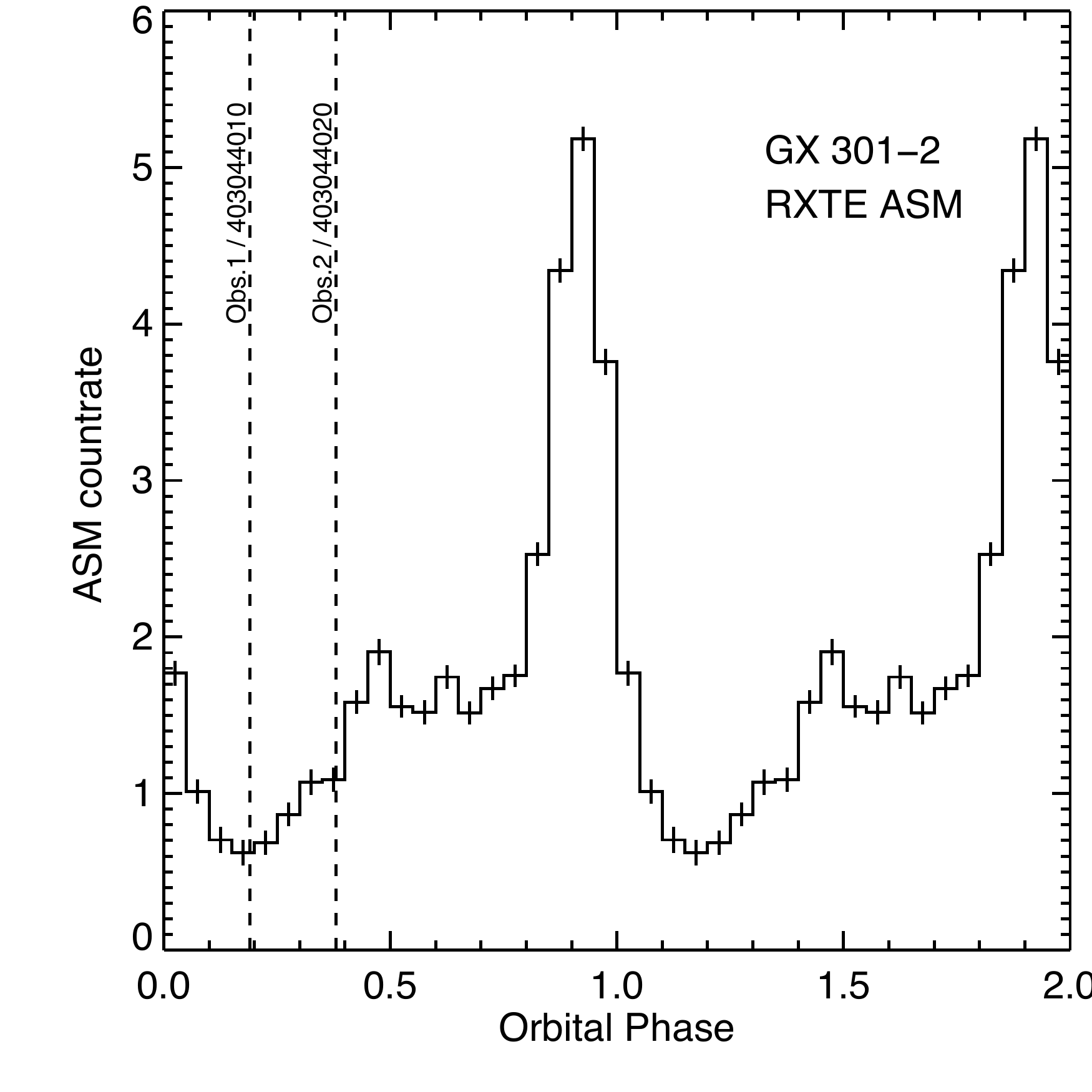}
\caption{\textbf{Left:} \textsl{Swift}-BAT 15--50\,keV light curve of
  the transient accreting pulsar 1A\,1118$-$61 showing the 2009
  outburst of the source. Vertical lines indicate the times of the
  January 15, 2009, and January 28, 2009, \textsl{Suzaku} pointings
  (exposures of $\sim$50\,ks and $\sim$30\,ks,
  respectively). \textbf{Right:} \textsl{RXTE}-ASM 1--12\,keV orbital
  profile of the persistent accreting pulsar GX\,301$-$2.  Vertical
  lines indicate the orbital phases of the August 25, 2008, and
  January 5, 2009, \textsl{Suzaku} pointings (exposures of
  $\sim$10\,ks and $\sim$60\,ks, respectively). After
  \citep{suchy:11a} and \citep{suchy:11b}.}
\label{fig:lc}
\end{figure}

Accreting X-ray pulsars were first discovered in the 1970s with the
detection of the 4.8\,s pulsation of \mbox{Cen X-3}
\citep{giacconi:71a}. In these sources the compact object is a highly
magnetized neutron star leading to accretion of material along the
magnetic field lines and the formation of X-ray emitting accretion
columns above the neutron star's magnetic poles. These sources can
also show absorption-line-like scattering features in their X-ray
spectra at multiples of $E_\mathrm{cycl}\sim11.6\,\mathrm{keV}\times B
[10^{12}\,\mathrm{G}]$ since for such strong magnetic fields electron
energies are quantized. Observations of sources with cyclotron
resonant scattering features (CRSF, cyclotron lines) provide us with
the most direct information for studying strong magnetic
fields. \textsl{Suzaku} is especially well suited to find/study
cyclotron lines due to its broadband X-ray coverage at high
sensitivity with comparatively low background. With O- or B-type donor
stars most CRSF sources are high mass X-ray binaries (exceptions: Her
X-1, 4U\,1626$-$67). Those in the first block of
Table~\ref{tab:sources} (plus persistent X Per and some of the
potential CRSF candidates in the third block) are Be type X-ray
binaries which are generally transient \citep{reig:11a}. They
typically have eccentric orbits and their X-ray activity is triggered
when the neutron star crosses a disk of material that surrounds the
equator of the Be/Oe star. Between outbursts they can show years long
quiescence periods. While the CRSF sources listed in the second block
of Table~\ref{tab:sources} are persistent X-ray emitters they are very
variable on all observed time scales, reflecting stellar wind,
magnetosphere (see, e.g., \citep{fuerst:10a,doroshenko:10a} for
Vela~X-1), or accretion disk structure (see, e.g.,
\citep{staubert:09a} for Her~X-1). Fig.~\ref{fig:lc} shows examples of
typical long term variability for a transient and a persistent
source. For recent reviews on accreting pulsars see
\citep{caballero:11b} and \citep{paul:11a}.

\begin{table}
\begin{tabular}{rllcclc}
\hline
  & \tablehead{1}{l}{t}{Source Name}
  & \tablehead{1}{l}{t}{CRSF \\ Energies \\ $E_\mathrm{cycl}$ [keV]}
  & \tablehead{1}{l}{t}{Pulse\\ Period \\ $P_\mathrm{pulse}$ [s]}
  & \tablehead{1}{l}{t}{Orbital\\ Period \\ $P_\mathrm{orbit}$ [d]}
  & \tablehead{1}{l}{t}{Type}
     & \tablehead{1}{l}{t}{\textsl{Suzaku}\\ analysis \\ reference}\\
\hline
1 & \textit{Swift\,J1626.6$-$5156}  & 10                 &  15           & 132.9  & T, Be           & \\
2 & 4U\,0115$+$63                   & 14, 24, 36         & 3.6           & 24.31  & T, B0.2 Ve      & work in progress\\
  &                                 & 48, 62             &               &        &                 & \\
3 & \textit{V\,0332$+$53}           & 27, 51, 74         & 4.37          & 34.25  & T, O8.5 Ve      & \\
4 & \textit{Cep~X-4}                & 28                 & 66.25         & >23    & T, B1.5 Ve      & \\
5 & \textit{MXB\,0656$-$072}        & 33                 & 160           & ?      & T, O9.7 Ve      & \\
6 & XTE\,J1946$+$274                & 36                 & 15.8          & 169.2  & T, B0-1 V-IVe   & work in progress\\
7 & 1A\,0535$+$26\tablenote{Caballero et al., see elsewhere in these proceedings} & 45, 100 & 105 & 110.58 & T, O9.7 IIe & \citep{terada:06a}\\
  &                                 &                    &               &        &                 & \citep{naik:08a}\\
  &                                 &                    &               &        &                 & \citep{caballero:11a}\\
8 & GX\,304$-$1\tablenote{Yamamoto et al., see elsewhere in these proceedings} & 54 & 272 & ? & T, B2 Vne & \citep{yamamoto:11a}\\
9 & \textbf{1A\,1118$-$616}         & 55, 112?           & 408           & 24     & T, O9.5 IV-Ve   & \citep{suchy:11a}\\         
\hline\\
10 & \textbf{4U\,1907$+$09}         & 19, 40             & 438           & 8.38   & P, B2 III--IV   & \citep{rivers:10a}\\
11 & \textit{4U\,1538$-$52}         & 20                 & 530           & 3.73   & P, B0 I         & \\
12 & \textbf{Vela~X-1}              & 25, 53             & 283           & 8.96   & P, B0.5 Ib      & \citep{doroshenko:11a}\\
13 & \textit{X~Per}                 & 29                 & 837           & 250.3  & P, B0 III--Ve   & \\
14 & Cen~X-3                        & 30                 & 4.8           & 2.09   & P, O6.5 II      & \citep{naik:11a}\\
15 & \textbf{GX\,301$-$2}           & 37                 & 690           & 41.5   & P, B1.2 Ia      & \citep{suchy:11b}\\
16 & 4U\,1626$-$67                  & 37                 & 7.66          & 0.028  & P, LMXB         & \citep{camero:11a}\\
17 & Her~X-1                        & 39                 & 1.24          & 1.7    & P, A9-B         & \citep{enoto:08a}\\
   &                                &                    &               &        &                 & \citep{klochkov:11a}\\
\hline\\
18 & EXO 2030$+$375                 & 11? 63?            & 42            & 46.0   & T, B0 Ve        & work in progress\\
19 & GRO J1008$-$57                 & 88?                & 93.7          & 249.5  & T, B0e          & \citep{naik:11b}\\
   &                                &                    &               &        &                 & \citep{kuehnel:11a}\\
20 & \textit{LS V $+$44 17}         &                    & 202           &  ?     & T, B0.2 Ve      & \\
21 & \textit{GS\,1843$+$00}         & 20?                & 29.5          &  ?     & T, B0-2 IV-Ve   & \\
22 & XTE J1739$-$302\tablenote{Bodaghee et al., see elsewhere in these proceedings} & & ? & 51.47  & T, O8 Iab & \citep{bodaghee:11a}\\
23 & \textit{OAO 1657$-$415}        & 36?                & 37.7          & 10.4   & P, B0-B6 Ia-Iab & \\
24 & 4U 1700$-$377                  & 37?                & ?             & 3.4    & P, O6.5 Iaf+    & work in progress\\  
25 & LMC X-4                        & 100?               & 13.5          & 1.41   & P, O7 IV        & \citep{hung:10a}\\
26 & 4U 1909$+$07                   &                    & 604           & 4.4    & P, OB           & work in progress\\
27 & IGR\,16318$-$4848              &                    & ?             & ?      & P, sgB[e]       & \citep{barragan:09a}\\
\hline
\end{tabular}
\caption{Properties of a sample of accreting pulsars. 1--9: All known
  transient CRSF sources. 10--17: All known persistent CRSF
  sources. 18--27: Selected potential CRSF candidates. All sources
  with names not printed in italics have been observed with
  \textsl{Suzaku} ($1\times-4\times$) at the time of writing. The
  pulse period $P_\mathrm{pulse}$ usually varies with luminosity and
  time. The same is often true for the cyclotron line energies
  $E_\mathrm{cycl}$ which can also vary with pulse phase. Here we
  quote representative values from the overall
  literature. \textbf{\textsl{Suzaku} results for sources with names
    in bold face are discussed in this work.}}
\label{tab:sources}
\end{table}

\section{\textsl{Suzaku} Observations}

\subsection{Discovery of a CRSF in 1A\,1118$+$61}

\begin{figure}
\includegraphics[width=0.49\textwidth]{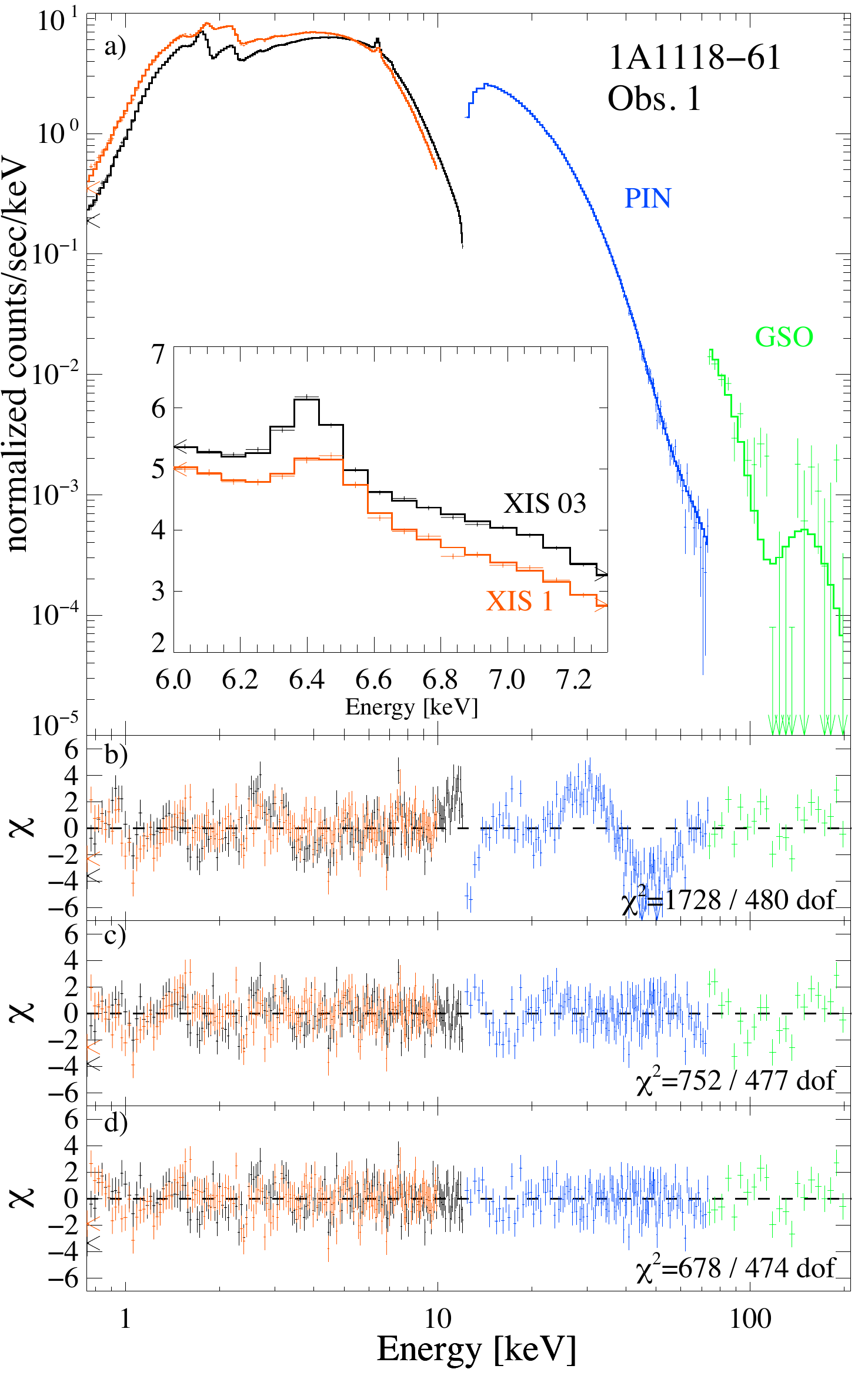}
\includegraphics[width=0.49\textwidth]{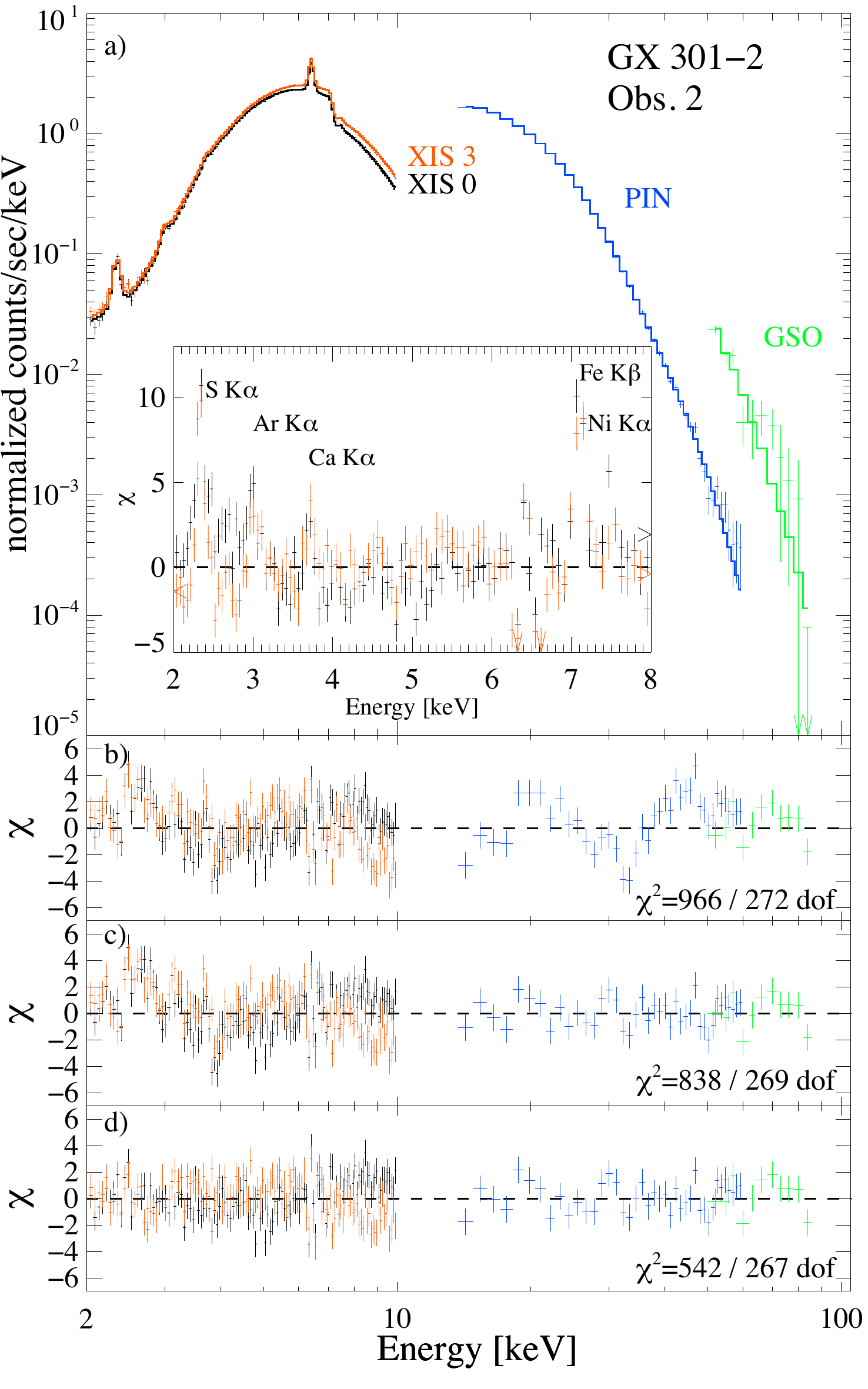}
\caption{\textbf{Left:} The main panel shows the \textsl{Suzaku} broad
  band spectrum and best fit of the brighter 1A\,1118$-$61 observation
  with an inset highlighting the iron line region. Residuals are shown
  including no CRSF (a), including one CRSF at 58\,keV (b), and
  including a second CRSF at 112\,keV (c, best
  fit). \textbf{Right:}. The main panel shows the \textsl{Suzaku}
  broad band spectrum and best fit of the brighter GX\,301$-$2
  observation with an inset highlighting the emission line residuals
  from neutral material which remain after modeling Fe K$\alpha$. (The
  back illuminated XIS1 was modeled separately here, with consistent
  results.) Residuals are shown including no CRSF and not accounting
  for partial covering absorption (a), including a CRSF at 35\,keV
  (b), and also accounting for partial covering (c, best fit). After
  \citep{suchy:11a} and \citep{suchy:11b}.}
\label{fig:spec}
\end{figure}

After only two previously observed outbursts (1974, 1992) the Be X-ray
binary 1A\,1118$-$61 went into outburst in January 2009 and was
observed with \textsl{Suzaku} at the $\sim500$\,mCrab outburst peak
and two weeks later at the end of the main outburst
(Fig.~\ref{fig:lc}, left). Together with quasi-simultaneous
\textsl{RXTE} observations \citep{doroshenko:10a} these observations
enabled the discovery of a $\sim55$\,keV cyclotron line in this source
\citep{suchy:11a} (Fig.~\ref{fig:spec}, left). \textsl{RXTE},
\textsl{Swift}-BAT, and \textsl{Suzaku} also co-discovered another
cyclotron line, namely the $\sim54$\,keV feature in GX\,304$-$1
\citep{sakamoto:10a,yamamoto:11a}. Furthermore there are indications
that the line in 1A\,1118$-$61 may be luminosity dependent. With
$E_\mathrm{cycl}$ values of $58.2^{+0.8}_{-0.4}$\,keV and
$47.4^{+3.2}_{-2.3}$\,keV for the flare and decline observation,
respectively, the source would show a positive $L$-$E_\mathrm{cycl}$
correlation like Her~X-1. Negative or no correlations have recently
been observed for other CRSF sources \citep{caballero:11b}. A possible
explanation for the existence of positive and negative correlations is
the different state of the accretion column in the sub- and
super-Eddington regimes \citep{staubert:07a}.

\subsection{Pulse Phase Resolved Analysis of GX\,301$-$2}

\begin{figure}
\includegraphics[width=0.49\textwidth]{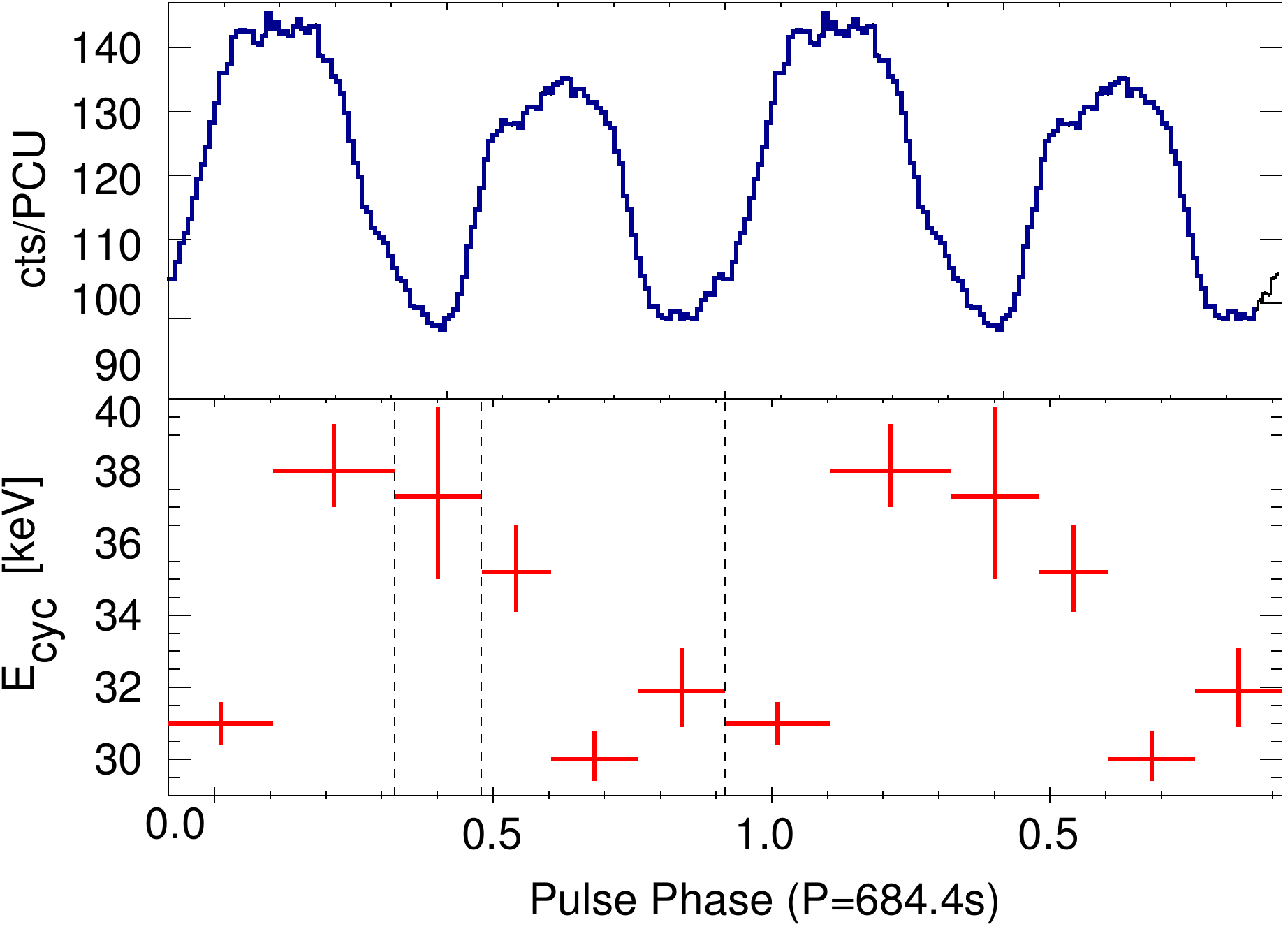}
\includegraphics[width=0.49\textwidth]{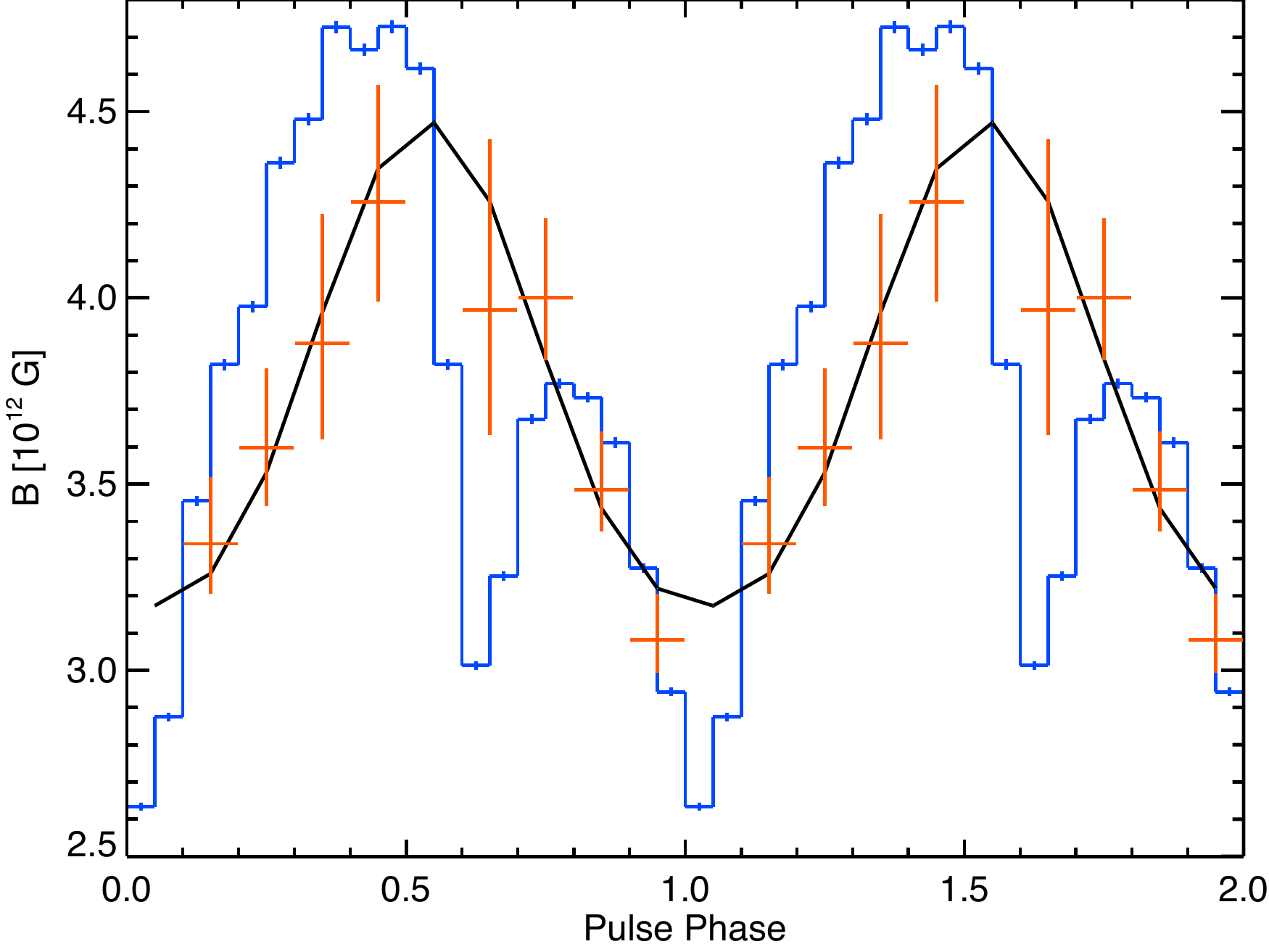}
\caption{\textbf{Left:} Results from a $\sim200$\,ks \textsl{RXTE}
  observation of GX\,301$-$2, after \citep{kreykenbohm:04a}. The upper
  panel shows the double peaked pulse profile in the 19-60\,keV energy
  band (twice for clarity). The lower panel shows that the cyclotron
  line energy varies by $\sim$20\% over one cycle. Vertical lines
  indicate pulse profile minima. \textbf{Right:} Results from a
  $\sim60$\,ks \textsl{Suzaku} observation of GX\,301$-$2 (``obs 2''
  from Fig.~\ref{fig:lc} and Fig.~\ref{fig:spec}), after
  \citep{suchy:11b}. The red data points confirm the $E_\mathrm{cycl}$
  variations, translated into directly $E_\mathrm{cycl}$ proportional
  $B$-field variations. The black curve shows the best fit magnetic
  dipole model described in the text. The blue histogram indicates
  the 1--10\,keV pulse profile (arbitrary normalization).}
\label{fig:dipole}
\end{figure}

The persistent accreting pulsar GX\,301$-$2 was observed twice with
\textsl{Suzaku}, during less bright and less well studied orbital
phases outside of the pre-periastron flare (Fig.~\ref{fig:lc},
right). These observations allowed for a detailed study of the highly
absorbed phase averaged spectrum, including several fluorescence
emission lines and the well known 37\,keV CRSF (Fig.~\ref{fig:spec},
right) as well as of pulse phase resolved spectra for the second,
longer observation \citep{suchy:11b}. Strong variations of the
cyclotron line energy $E_\mathrm{cycl}$ with pulse phase had been
discovered in an \textsl{RXTE} observation dominated by the
comparatively bright pre-periastron flare \cite{kreykenbohm:04a}
(Fig.~\ref{fig:dipole}, left, lower panel). Several other CRSF sources
are known to show similarly strong variations (see, e.g.,
\citep{suchy:08a} and references given by
\citep{kreykenbohm:04a,suchy:11a}). Even though the \textsl{Suzaku}
observation does not cover the bright pre-periastron flare a clear
pattern for the $E_\mathrm{cycl}$ variations emerges within a
comparatively modest exposure time (Fig.~\ref{fig:dipole},
right). Assuming a simple magnetic dipole the $B$-field projected onto
the line of sight can be calculated for each phase bin, depending on a
set of geometric angles. This leads to four symmetrically equivalent
best fit solutions. Model constraints derived using the observed ratio
of the width of the cyclotron line and $E_\mathrm{cycl}$ favor a
solution wherein the spin axis is tilted by $15^{\circ}$ with respect
to the line of sight \citep{suchy:11b}.

\subsection{Dips \& Flares: Stellar Wind Structure}

\begin{figure}
\includegraphics[width=0.49\textwidth]{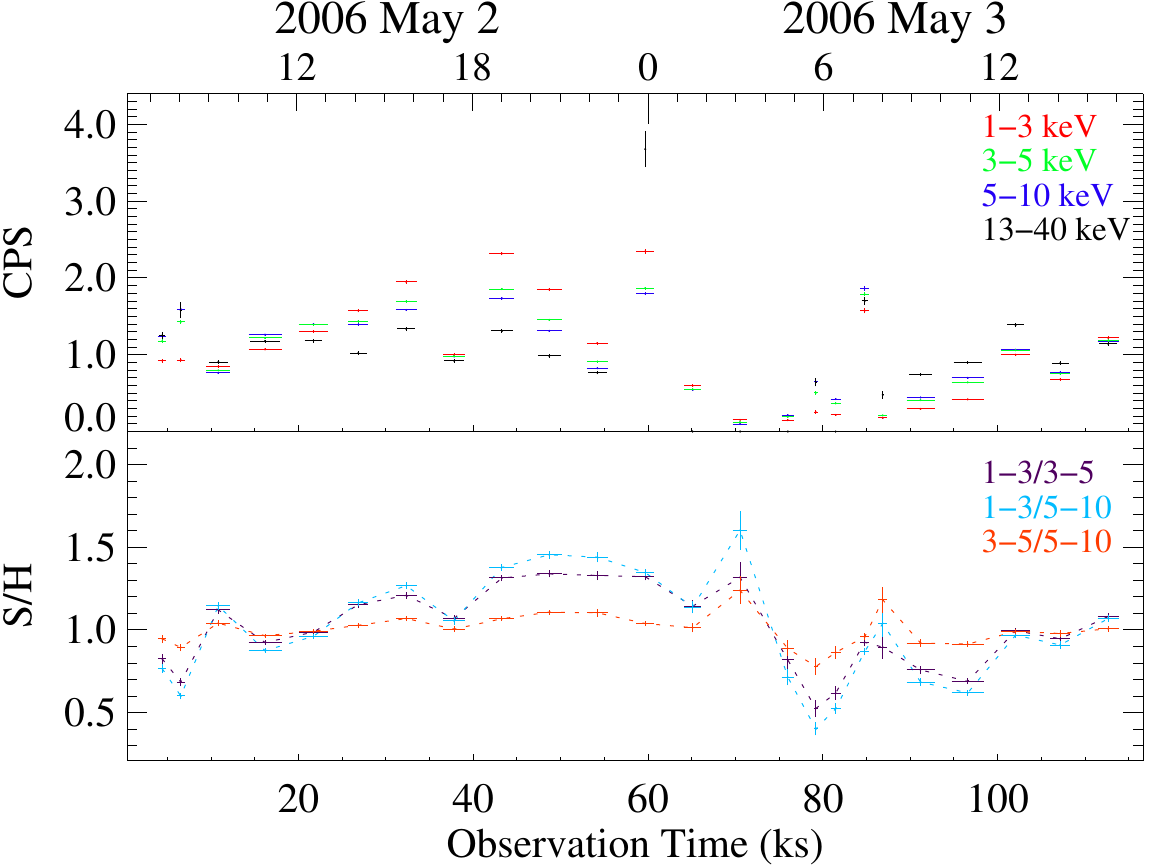}
\includegraphics[width=0.49\textwidth]{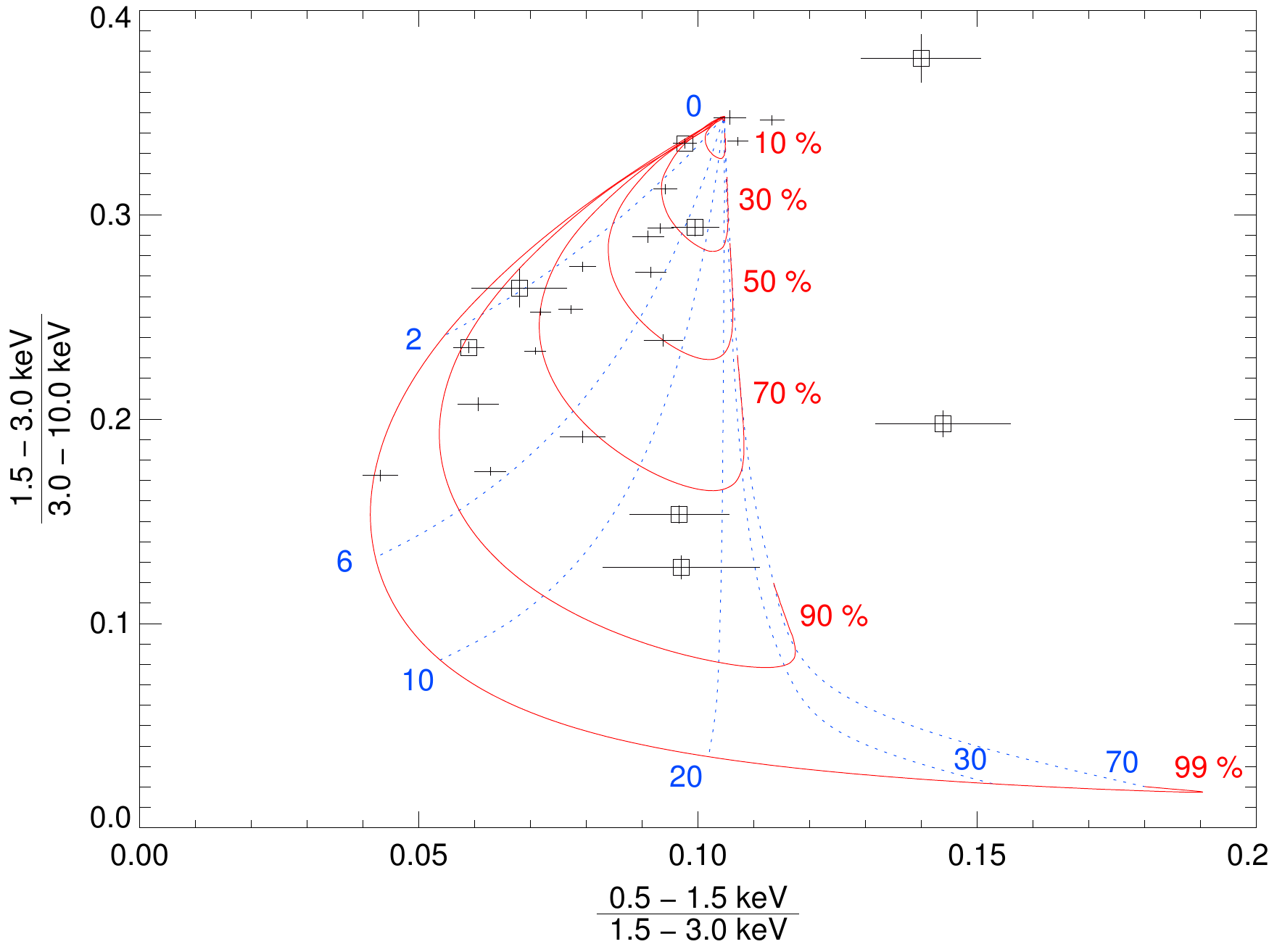}
\caption{\textbf{Left:} Energy-resolved, background-subtracted light
  curves (upper panel), and softness ratios (lower panel) from XIS0
  for the 2006 \textsl{Suzaku} observation of
  4U\,1907$+$09. \textbf{Right:} Color-color diagram from added XIS0
  and XIS3 data for the same observation. Boxed data points correspond
  to the deep dip in the light curve between 60\,ks and 100\,ks. Grid
  lines indicate hardness values from spectral models obtained by
  increasing the covering fraction at constant $N_\mathrm{H}$ (blue)
  or vice versa (red) compared to the best fit spectral model for
  times without additional partial covering absorption or
  flaring. From \citep{rivers:10a}.}
\label{fig:1907}
\end{figure}

\subsubsection{4U\,1907$+$09}

The $\sim$20\,mCrab accreting pulsar 4U\,1907$+$09 is characterized by
a highly variable persistent flux as well, but in this case the dips
and flares do not seem to be related to the orbital phase
\citep{intzand:97a,rivers:10a}. \textsl{Suzaku} observed the source on
May 2--3, 2006, and April 19--20, 2007, for respectively $\sim$60\,ks
and $\sim80$\,ks. Both observations display dips and flares
(Fig.~\ref{fig:1907}, left, shows the 2006 case). It had been noted
before that the dipping might not be due to absorption but due to
cessation of accretion \citep{intzand:97a}. Using color-color diagrams
\citep{rivers:10a} found that while some of the flux and hardness
variability is consistent with absorption there are indeed events
where this is not the case, i.e., for the deep dip in the 2006 dataset
(Fig.~\ref{fig:1907}, right). They conclude that the overall behavior
is consistent with a clumpy wind.

\subsubsection{Vela X-1}
Even more extreme flux variability, so called off states and giant
flares -- both with flux changes of up to a factor of $\sim$20 -- have
been discovered with \textsl{INTEGRAL} for Vela X-1
\citep{kreykenbohm:08a}. During the off states no pulsations were
observed with \textsl{INTEGRAL}. \textsl{Suzaku} detected three
off-sates during a 100\,ks observation on June 17--18, 2008, lasting
up to 3.2\,ks \citep{doroshenko:10a}. Due to \textsl{Suzaku's}
sensitivity pulsations could still be detected. \citep{doroshenko:10a}
suggest that while the off states might be triggered by stellar wind
inhomogeneities they are consistent with gated accretion where
Kelvin-Helmholz instabilities allow for some matter to leak into the
magnetosphere.

\section{A Physically Motivated CRSF Model}

\begin{figure}
\includegraphics[width=0.95\textwidth]{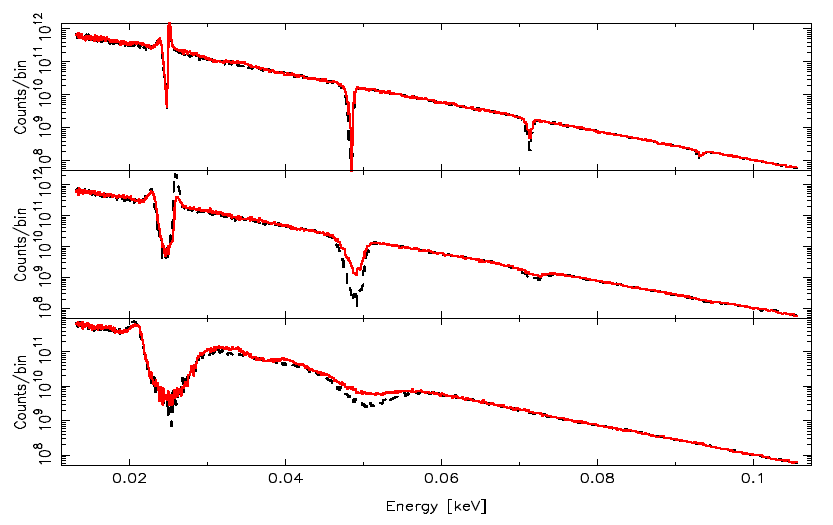}
\caption{Comparison of spectra resulting from the previous (black) and
  current (red) version of the MC code for simulating physical CRSFs,
  for three different photon incident angles. From
  \citep{schwarm:10a}. The input X-ray continuum spectrum is an
  empirical one (a power law with a Fermi-Dirac cutoff, characterized
  by $E_\mathrm{fold}=10$\,keV and $E_\mathrm{cut}=12$\,keV). The
  lines, however, are produced in a cylindrical accretion column with
  a magnetic field of $B/B_\mathrm{crit}=0.05$ and an electron plasma
  temperature and optical depth of $kT_\mathrm{e}=3$\,keV and
  $\tau_\mathrm{es}=10^{-3}$, respectively. Note that
  $B_\mathrm{crit}=m_\mathrm{e}^2c^3/e\hbar=4.414\times10^{13}$\,G.}
\label{fig:mc}
\end{figure}

Currently empirical models are used to describe cyclotron line
features, parametrized, e.g., as absorption with a Gaussian optical
depth profile or by a Lorentzian profile (\texttt{gabs} and
\texttt{cyclabs} in \texttt{xspec}, respectively). Work has been going
on since the 1990s to describe CRSFs in terms of physical parameters
like the temperature $kT_\mathrm{e}$ and optical depth
$\tau_\mathrm{es}$ of the electron plasma in the line forming region,
as well as the $B$-field strength and the accretion geometry
\citep{harding:91a,araya:99a}. This endeavor reached a recent point of
culmination with the implementation of a Monte Carlo (MC) code
describing the scattering process, the calculation of a grid of
Green's functions which can be used to apply the process to any
continuum spectrum, and the development of the \texttt{xspec} model
\texttt{cyclomc} for this task by \citep{schoenherr:07a}. First
\texttt{cyclomc} applications to observed X-ray spectra, e.g., of
V\,0332$+$53, resulted in generally acceptable fits and parameter
values, with the caveat of stronger emission wings than observed and
different relative line depths than expected
\citep{schoenherr:07a}. With a correction of the scattering cross
sections in the most recent iteration of the MC code by
\citep{schwarm:10a} this has been resolved. Fig.~\ref{fig:mc} shows a
comparison of CRSFs calculated with the two MC versions for three
different photon incident angles with respect to the magnetic field in
a cylindrical accretion column. The team is currently in the process
of calculating a new grid of Green's functions to be used with
\texttt{cyclomc}.

\section{Summary and Conclusions}

\vspace*{-0.5\baselineskip}

So far \textsl{Suzaku} has observed eleven of the seventeen cyclotron
line sources at least once. Two were identified as CRSF sources for
the first time (1A\,1118$-$61, GX\,304$-$1). Detailed phase resolved
studies can be performed that allow us to constrain the $B$-field
geometry (e.g., GX 301-2). The important $E_\mathrm{cycl}-L$
relationship can be studied (e.g., 1A\,1118$-$61, 1A\,0535$+$26),
potentially with much better time resolution than before (future
monitoring observations are desirable). Detailed studies of the mass
transfer process (stellar wind, magnetosphere) can be performed as
well (e.g., 4U\,1907+09, Vela~X-1, Cen X-3). Future work is foreseen
to include a comparative study of the \textsl{Suzaku} sample applying
the new \texttt{cyclomc} line model.


\vspace*{0.3\baselineskip}

\footnotesize{\textbf{Acknowledgments}: We thank the \textsl{Suzaku}
  team for their help in executing these observations, especially for
  the excellent timing of the 1A\,1118$-$61 ToO (Fig.~\ref{fig:lc},
  left) and the eclipse-to-eclipse Cen~X-3 observation (not discussed
  here due to space limitations).}

\vspace*{-1.2\baselineskip}


\end{document}